\documentclass[preprint]{aastex631}
\received{ }
\revised{ }
\accepted{ }

\shorttitle{Helioseismic Investigation of Quasi-biennial Oscillation Source Regions}
\shortauthors{Jain, Chowdhury \& Tripathy}

\begin{document}

\title{Helioseismic Investigation of Quasi-biennial Oscillation Source Regions}

\correspondingauthor{Kiran Jain}
\email{kjain@nso.edu}

\author[0000-0002-1905-1639]{Kiran Jain}
\affiliation{National Solar Observatory, Boulder, CO 80303, USA}

\author[0000-0003-3253-9054]{Partha Chowdhury}
\affiliation{University College of Science and Technology, Department of Chemical Technology, \\University of Calcutta, Kolkata, 700009, West Bengal, India}

\author[0000-0002-4995-6180]{Sushanta C. Tripathy}
\affiliation{National Solar Observatory, Boulder, CO 80303, USA}



\begin{abstract}
 We studied the temporal evolution of quasi-biennial oscillations (QBOs)  
using acoustic mode oscillation frequencies from the Global Oscillation 
Network Group.  The data used here span over more than 25 yr, covering 
solar cycles 23 and 24 and the ascending phase of cycle 25.  The analysis
  reveals  that the QBO-like signals are present in both the cycles,
 but with different periods.  The dominant QBO period in cycle 23 is found 
to be about 2 yr while it is about 3 yr in cycle 24.  Furthermore, the 
quasi-biennial oscillatory signals are present only during the ascending 
and high-activity phases of cycle 23 and quickly weaken around 2005 
during the declining phase. In comparison, the QBO signals are present 
 throughout the cycle 24, starting from 2009 to 2017.  We also explored 
the depth dependence in QBO signals and obtained a  close agreement at
 all depths, except in the near-surface shear layer. A detailed analysis 
of the near-surface shear layer suggests that the 
source region of QBOs is probably
 within a few thousand kilometers just below the surface.

\end{abstract}



\section{Introduction} \label{s-intro}

Solar activity  displays a variety of periodicities; some are long term and others are short term \citep{Kollath09}. Records of sunspot numbers\footnote{\url{https://www.sidc.be/silso/datafiles}} over multiple centuries reveal both types of periodicities. Among them, the  most dominant cyclic behavior is commonly known as the solar  activity cycle or Schwabe cycle \citep{Schwabe1844}, the period of which ranges from 9 to 13 yr. It is present in all activity measures observed in different layers of the solar atmosphere. Since the consistent and uninterrupted observations of 5 minute acoustic modes  \citep{Leighton62} are also available for more than two solar cycles, it is now possible to  probe the variability and structure of the solar interior in great detail \citep{JCD02}. Similar to above-surface activity indicators, strong 11 yr cyclic patterns are also found in the change of helioseismic {\it p}-mode frequencies computed using the methods of global helioseismology \citep[][ and references therein]{Jain22a} as well as local helioseismology \citep{Tripathy13, Tripathy15}. These changes are strongly correlated with the variations in solar magnetic activity, though the correlation between them differs depending on the phase of the cycle \citep{Jain09}. The long time series have further allowed us to uncover several new features in helioseismic data that were inaccessible otherwise, e.g., recently discovered high-latitude inertial modes \citep{Gizon21}. 

In addition to 11 yr cyclic patterns, other periods have also been identified in both solar activity proxies \citep[e.g.,][]{Ulrich13,Chowdhury19} and  helioseismic data obtained from different instruments \citep[e.g.,][]{howe00, Broomhall09}.  During the early years of the operation of the Michelson Doppler Imager (MDI) on board the Solar and Heliophysics Observatory (SoHO), the oscillatory component of a 1 yr  period was reported in the oscillation frequencies \citep{Antia01, Jain03}, which was later found to be an artifact in the data, as it matched with the orbital period of the Earth. However, shorter quasiperiodic variations or quasi-biennial oscillations (QBOs)  with periods ranging from 0.6 to 4 yr  exist in all types of data \citep[][, and references therein]{Bazilevskaya14}. These quasiperiodic variations have also been identified in several energetic events, e.g., solar flares \citep{Kilcik20}, coronal mass ejections \citep{Li23}, as well as meteorological parameters \citep{Anstey22}. Studies based on solar observations show that the amplitude of the QBOs varies with the activity cycle, with the highest amplitudes during solar maxima that become weaker during the activity  minima. During most solar cycle maxima, double peaks (Gnevyshev Gap) are more prevalent in activity indices, due to the asymmetry between the northern and southern hemispheres \citep{Norton10,Ravindra22}, where one hemisphere achieves the maximum amplitude earlier than the other hemisphere. It is also suggested that the solar cycle may evolve independently in the two hemispheres \citep[e.g.,][]{Veronig21}, which introduces asymmetry in both hemispheres.  This asymmetry is also believed to be linked with the higher amplitude of the QBOs at solar maxima, especially in global indices. Thus,  all these studies suggest the existence of  two magnetic cycles with different periodicities that have been discussed by several authors \citep{Benevolenskaya98, Vecchio08}.

The magnetic field is believed to be generated in convection zone and then transported upward to different  layers of the Sun's atmosphere. During this process, the magnetic field slowly gets dissipated and a part of it reaches up to interplanetary space. Thus, the periodic variations obtained at different layers seem to be connected. In addition, the availability of asteroseismic observations of thousands of stars from Kepler and CoROT provide evidence of stellar magnetic cycles \citep[e.g.,][]{Garcia10b, Santos23}.  Therefore, a  precise understanding of different magnetic cycles progressing simultaneously is crucial for the better understanding of solar and stellar dynamos, and their variability.   \citet{Dikpati17} argued that the  quasiperiodic energy exchange among magnetic fields, Rossby waves, and differential rotation of the solar interior  are important to explain the explosive and quiet periods during the activity cycle.  Furthermore, the quasi-biennial pulsations observed in both solar and stellar flares suggest the similar underlying physics of these flares \citep{Broomhall19}. However, the stellar data are currently limited by the length of the observations. Therefore, the knowledge of solar dynamos can be expanded to understand the stellar dynamics. 

To get a deeper understanding of QBOs and their origin,  we present an analysis of the QBO periodicities observed in the last two solar cycles by studying acoustic mode frequencies. Note that cycles 23 and 24 had very different amplitudes and were separated by an extremely long period of very low solar activity. Thus, the similarities and differences in QBOs in these cycles are important in constraining solar dynamo models. The {\it p}-mode frequency data and the method are described in Section~2. We present the results in Section~3 and the possible origin of QBOs is discussed in Section 4. Finally, we summarize our findings in Section 5.

\begin{figure}[t]
\begin{center}
\includegraphics[width=0.54\textwidth,  height=14.5cm,angle=90]{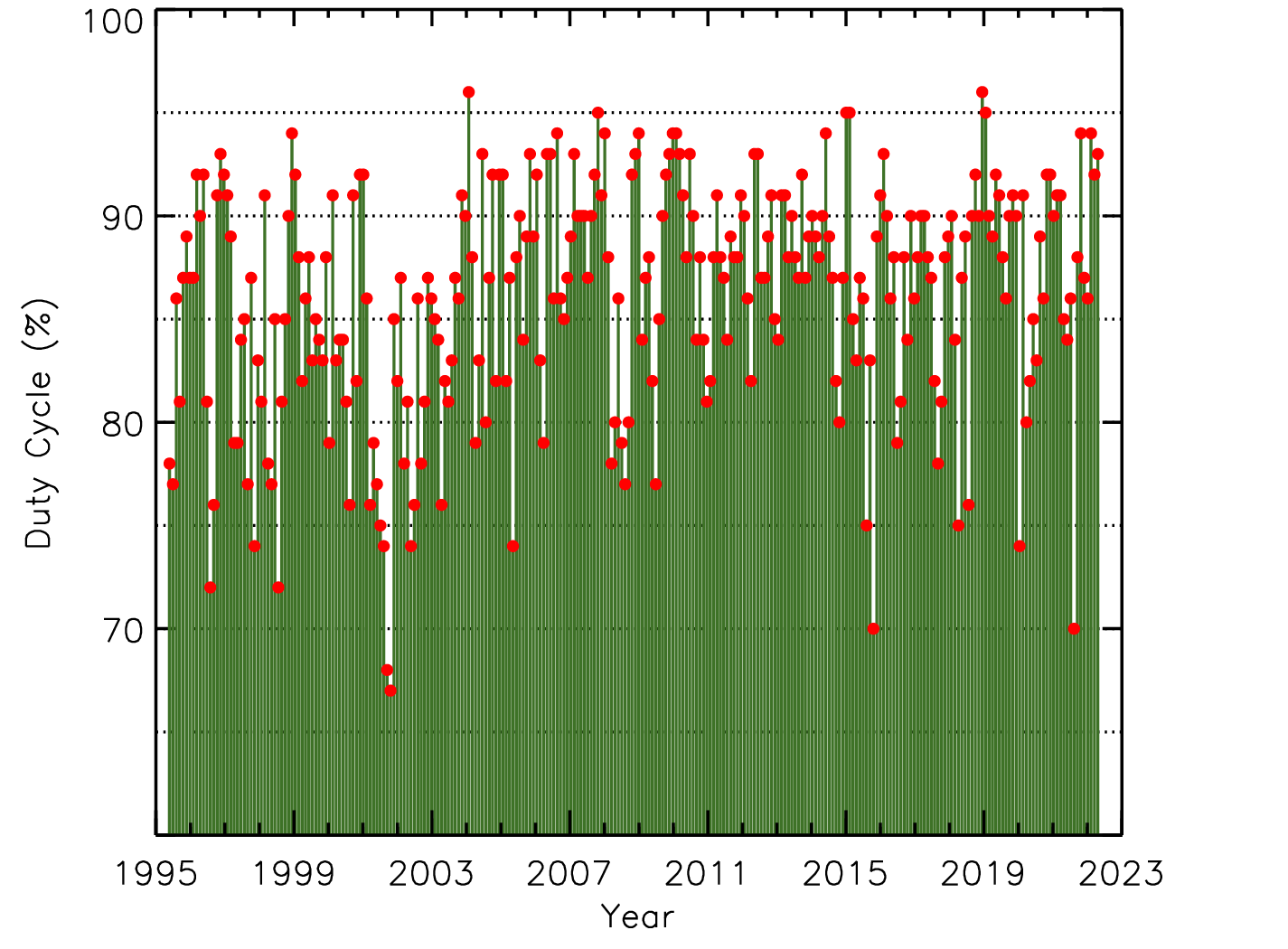}  
\caption{ Duty cycles of 36 day $p$-mode frequency data sets.  }
\label{DC}
\end{center}
\end{figure}
\section{Data and Technique} \label{sec-data}
 We utilized $p$-mode frequencies computed from Doppler observations of the Global Oscillation Network Group \citep[GONG;][]{harvey96}.  GONG is a ground-based network of six sites, and it has been providing unique and consistent frequency measurements with a significantly high duty cycle \citep{jain21a} for probing the solar interior for more than two solar cycles.  the data analyzed here consist of 274 nonoverlapping sets of 36 days, covering a period of 27 yr from 1995 May 7 to 2022 May 8 (two full solar cycles, 23 and 24). The duty cycles of these data are plotted in Figure~\ref{DC}, and the mean and median duty cycles are 86\% and 87\%, respectively.  The frequencies, $\nu_{n \ell m}$, were computed for the individual ($n, \ell, m$) multiplets,  where $n$ is the radial order, $\ell$ is the harmonic degree, and $m$ is the azimuthal order, running  from $-\ell$ to $+\ell$.  The mode frequency for each multiplet was estimated from the $m-\nu$ power spectra constructed from the time series of an individual 36 day period.  We used a GONG peak-fitting algorithm based on the multitaper spectral analysis coupled with a Fast Fourier Transform (FFT) to compute the power spectra \citep{Komm99}. Finally, we applied a minimization scheme guided by an initial guess table to the Lorentzian profiles to fit the peaks in the   $m-\nu$ spectra.
 The frequency and degree ranges covered in this work are  1860 $\le \nu \le$ 3450 $\mu$Hz and  0 $\le \ell \le$ 120, respectively.

\subsection{$p$-mode Frequency Shifts} \label{sec-helio}
\begin{figure}
\begin{center}
\includegraphics[width=0.74\textwidth,  height=13.5cm]{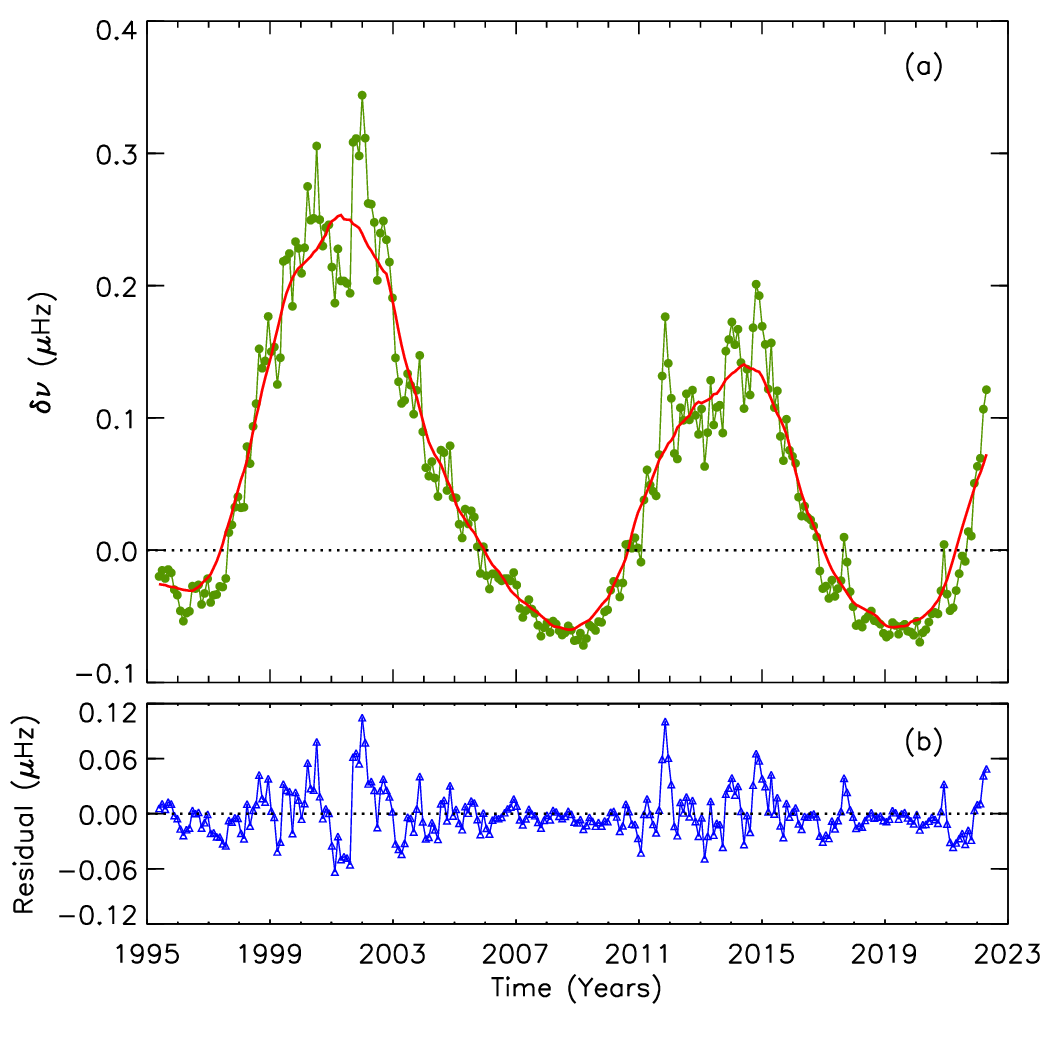}  
\caption{ Temporal evolution of: (a) the mean frequency shifts (green symbols connected with the solid line) and  the dominant 11 yr signal of the solar cycle as calculated by applying a boxcar filter of a width of 2 yr (solid red line):  and (b) the residual shifts after removing the 11 yr signal. The errors in frequency shifts are of the order of 10$^{-6}$ $\mu$Hz.
\label{dnu_all}}
\end{center}
\end{figure}

 To investigate the QBO periods in the acoustic oscillatory signal, we first calculated the change in frequencies with reference to the guess frequencies that are used for the fitting of $\nu_{n \ell m}$.  Since the frequency shifts have well-known dependencies on frequency and the mode inertia \citep[e.g.,][]{Jain00}, we scale the change in frequencies with mode inertia as described by \citet{JCD91}, while calculating the weighted mean frequency shift, $\delta\nu$, from the following relation:
\begin{equation}
\delta\nu(t) = \sum_{n \ell m}\frac{Q_{n \ell}}{\sigma_{n \ell m}^{2}}\delta\nu_{n \ell m}(t) / \sum_{n \ell m}\frac{Q_{n \ell}}{\sigma_{n \ell m}^{2}}
\label{eqn_nu}
\end{equation}
Here, $Q_{n \ell}$ is the inertia ratio, $\sigma_{n \ell m}$ is the uncertainty  in frequency determination, and $ \delta\nu_{n \ell m}(t)$ is the change in measured frequency for  a given $n$, $\ell$  and $m$.  We display in Figure~\ref{dnu_all}a  the temporal variation of $\delta\nu$ for the entire period. As illustrated,  the frequency shifts follow the trends of the solar activity cycles, confirming that the strength of cycle 24 was much weaker than the cycle 23. To extract the short-term fluctuations in $\delta\nu$, we follow the procedure as described by \citet{jain11} and subtract a smooth trend from the mean shifts by applying a boxcar filter with the width of 2 yr. The smoothed curve, shown in Figure~\ref{dnu_all}(a), depicts the 11 yr envelope of the solar activity cycle. Figure~\ref{dnu_all}(b) shows the residuals that are present in the oscillation frequencies  but not related with the 11 yr  activity cycle, and are believed to originate from the  short-term periodicities.

\subsection{ Morlet Wavelet Analysis} \label{sec-wavelet}
The wavelet analysis is a valuable tool for examining the presence of localized oscillations in the nonlinear time series, both in the time and frequency domains. Here 
we use the continuous wavelet transformation  of the Morlet wavelet tool to study the presence and temporal evolution of the QBOs in {\it p}-mode frequencies  
\citep{Torrence98}, 

\begin{equation}
\psi_{n}(\eta) = \pi^{-1/4}e^{i\omega_{0}n}e^{-\eta^{2}}/2.  
\label{eqn_wavelet}
\end{equation}
In this expression, $\omega_0$ is a nondimensional frequency and we have adopted $\omega_0 = 6 $ \citep{Torrence98,Chowdhury22}.   The thick dashed line in subsequent wavelet plots indicates the cone of influence, where the wavelet power reduces by a factor of $e^{-2}$ due to the edge effect, and the thin black contours indicate the periods above the 95\% confidence level, under a red-noise background \citep{Grinsted04}. We also compute the Global Wavelet Power Spectra (GWPS) by averaging over time at a given frequency. The 95\% confidence level of the GWPS plots is determined following the recipe of \citet{Torrence98}.


\begin{figure}
\begin{center}
\includegraphics[width=0.7\textwidth,  height=11.5cm]{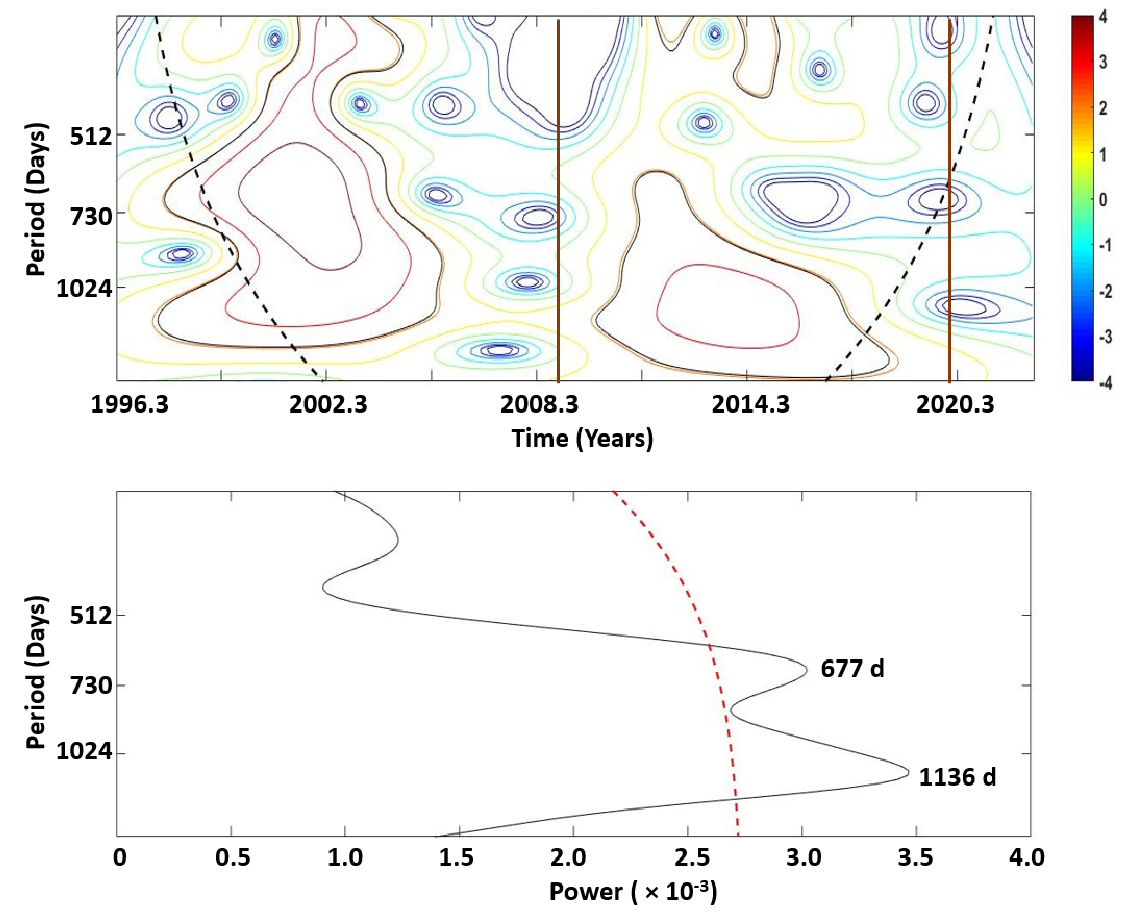}  
\caption{Morlet wavelet spectrum of the frequency shifts for all modes (top) and the global power spectra (bottom). The two vertical brown lines in the top panel delineate the boundaries between cycles 23 and 24 then  cycles 24 and 25,  while the dashed lines in both panels represent the 95\% confidence level.  
\label{QBO_all}}
\end{center}
\end{figure}

\begin{figure}
\begin{center}
\includegraphics[width=0.95\textwidth,  height=10.5cm]{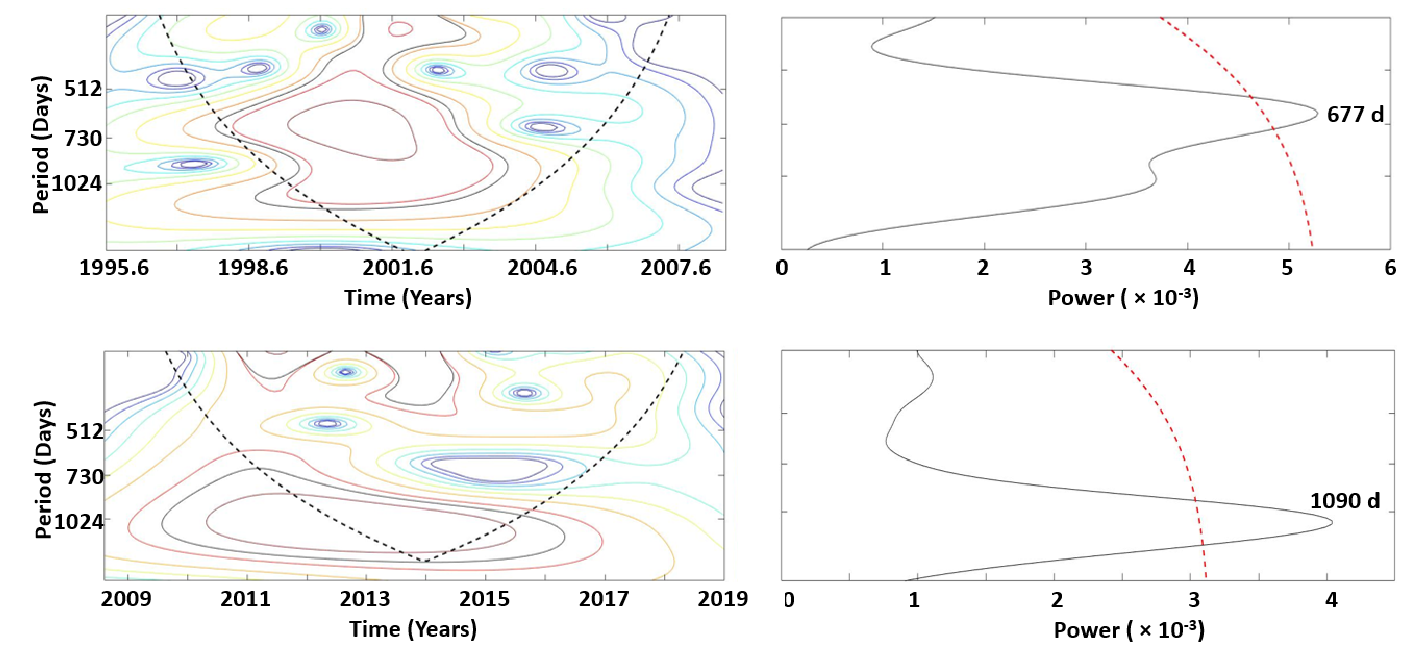}  
\caption{Morlet wavelet spectra (left) and the global power spectra (right) to study QBO periodicities for all modes modes in cycle 23 (top) and cycle 24 (bottom). The dashed lines represent the 95\% confidence level.  
\label{QBO_23_24}}
\end{center}
\end{figure}

\section{ Results} \label{sec-results}
 We calculated the weighted mean frequency shift over the frequency interval 1860$\mu$Hz$\le\nu\le$3450$\mu$Hz using Equation~\ref{eqn_nu}.  Since our aim is to study the periods relevant to QBOs, i.e., 1 -- 4 years, the Morlet wavelet spectra, displayed in Figure~\ref{QBO_all} and all subsequent figures, are limited to this period range.  As seen in Figure~\ref{QBO_all}, we obtain two distinct zones with significant power in  the wavelet spectrum; one during cycle 23 and the other during cycle 24. It is also seen that the QBO signal diminishes during the low-activity periods, while it  is enhanced during the high-activity periods. The variation in QBO power with time  is consistent with the previous studies based on various activity indices where a decrease in the QBO power was reported during the activity minimum by several authors \citep[][, and references therein]{Ravindra22}. The most striking feature in the wavelet spectrum is power distribution within the 95\% confidence level;  while it is confined to the  low-frequency part of the QBO spectrum with higher periods during 2009--2017, the spread is wider during 1997--2005 covering a large range of QBO periods.   In the GWPS displayed in the bottom panel of  Figure~\ref{QBO_all}, we  identify a prominent double-peak structure above the 95\% confidence level with QBO periods of 677$_{-82}^{+146} $ and 1136$_{-295}^{+133} $ days. These two QBO periods are primarily due to the differences in dominant power buildup in different parts of the spectrum.  It is important to mention that the double-peak structure  was not identified in earlier studies. For example, \citet{Tripathy13iau} reported only a single peak in the global power spectrum by using the oscillation frequencies for cycle 23 and the ascending phase of cycle 24.   Furthermore, the QBO power is found to disappear around 2005, during the declining phase of cycle 23, while a gradual decrease is seen throughout the declining phase of cycle 24 until 2017. The sudden loss of QBO power in the declining phase of cycle 23 might have resulted from the changes occurring in subsurface layers as reported by \citet{Basu12} and \citet{Howe17}. 

 To explore the origin of the two peaks in GWPS, shown in Figure~\ref{QBO_all}, we separated out the frequency shifts for cycles 23 and 24 and calculated the wavelet spectrum and GWPS for both cycles independently.    As is evident in Figure~\ref{QBO_23_24}, the double-peak structure no longer appears in the GWPS; only one peak is obtained in each cycle,  but with different QBO period and power; for  cycles 23 and 24,   the periods are 677$_{-69}^{+78} $ and 1090$_{-153}^{+206} $ days, respectively. These periods are consistent with those obtained from the analysis of the entire data, suggesting that both cycles have different dominant QBO periods. The QBO period obtained in cycle 23  confirms the single peak in the power spectrum as reported by \citet{Tripathy13iau} with a period of about 2 yr ($\sim$700 days).    We also obtain an insignificant hump around a period of  1010 days in cycle 23. Moreover, separate analyses for both cycles reveal that the QBO power in cycle 24 had weakened as compared to cycle 23.  We believe that the varying changes in subsurface layers during cycles 23 and 24 are responsible for the different periods.

\subsection{Depth dependence in QBOs } \label{depth}

Helioseismic studies focusing on the minima preceding cycles 23 and 24 reveal that the surface activity minimum occurred  about a year later than the minimum inferred in the deep solar interior, particularly in the radiative zone and core \citep{Salabert09, Tripathy10, Jain22a}. These results suggest the existence of more than one dynamo at different locations below the surface that may be affecting the variability in different zones. Thus, it is important to explore the influence of distinct zones on the QBO signals, if any.

For investigating the  QBO  periods for modes sensitive to  different zones below the surface, we divided the entire solar interior into three main zones: the core (0.0 $< r_t/R_{Sun} \le$0.3), the  radiative zone (0.3 $< r_t/R_{Sun} \le$ 0.7), and the convection zone (0.7 $< r_t/R_{Sun} \le$ 1.0).   The modes traveling to these zones can be restricted by using their ray path, defined by the lower- and upper-turning points,  where the lower-turning  point ($r_t$) determines the penetration depth  and the upper-tuning point provides information about the layer from which mode reflects back into the interior  \citep{JCD91}. The penetration depth  is calculated by the relation 
\begin{equation}
 r_t = \frac{c(r_t)}{2\pi}\frac{\sqrt{\ell(\ell+1)}}{\nu },
 \label{eqn_lower}
\end{equation}
where $c(r_t)$ is the sound speed at depth $r_t$. A higher value of $\nu/\sqrt{\ell(\ell+1)}$ denotes a smaller value of $r_t$ and hence a greater depth. Note that  all modes spend maximum time near the surface due to rapidly increasing sound speed with depth. Moreover, the upper-turning point defines a radius at which the acoustic cutoff frequency of the Sun (star) equals the frequency of the mode \citep[e.g., see Figure 5 of][]{Basu12}.  Modes with higher frequencies reach much closer to the surface.  Thus, we use the lower-and upper turning points defined by Equation~\ref{eqn_lower}  and the mode frequencies, respectively,  to infer the depth information. 

\begin{figure}
\begin{center}
\includegraphics[width=0.98\textwidth,  height=14.5cm]{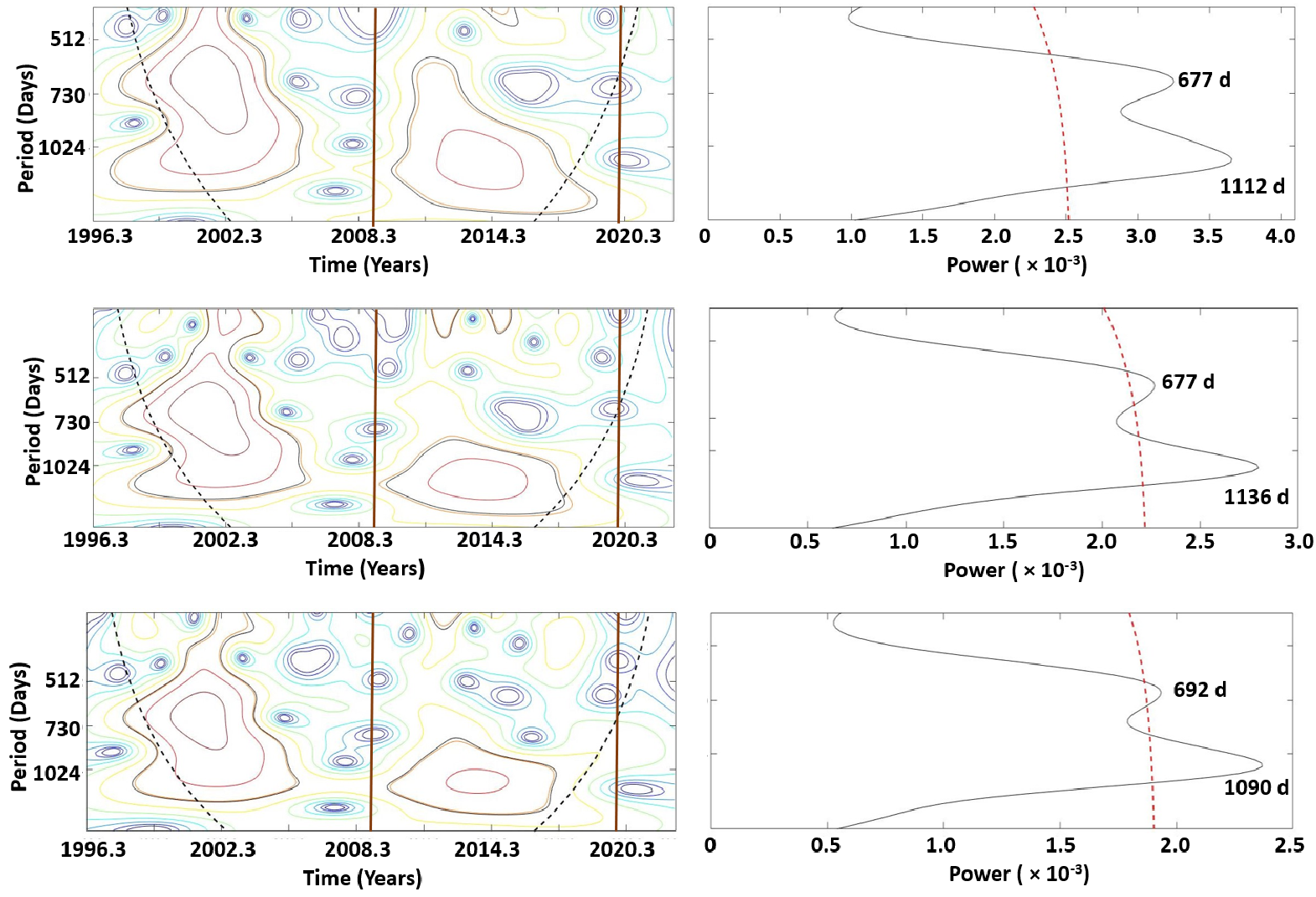}  
\caption{ Morlet wavelet spectrum of the frequency shifts (left) and the global power spectrum (right) for modes with lower-turning points in the convection zone (top), radiative zone (middle), and the core (bottom). The two vertical brown lines in the left panels delineate the boundaries between cycles 23 and 24 then cycles 24 and 25,  while the dashed lines in all panels represent the 95\% confidence level.  
\label{QBO_depth_dep}}
\end{center}
\end{figure}  

 The Morlet spectra and GWPS for modes with lower-turning points in  three different  zones are displayed in Figure~\ref{QBO_depth_dep}. It is worth mentioning that all {\it p}-modes irrespective of their lower-turning points travel to the surface, thus the frequencies are sensitive to the properties of the entire path (region) they travel through. For example, the modes returning to surface from the core will also reveal the conditions of the radiative and convection zones, while the modes returning from the convection zone will carry information from the convection zone only. As shown in Figure~\ref{QBO_depth_dep},  the QBO-type signals are present in different zones exhibiting similar double-peak structures. The obtained  QBO periods are:  692$_{-44}^{+63} $  and  1090$_{-173}^{+128} $ days for the modes returning from core: 677$_{-56}^{+78} $  and  1136$_{-139}^{+133} $  days for the modes returning from the radiative zone: and  677$_{-108}^{+146} $  and  1112$_{-271}^{+205} $ days for the modes confined to the convection zone only. There are some differences in the peak values, but they agree within the uncertainties.   A close agreement in the QBO periods implies no depth dependence in the QBOs.  These results are supported by a recent study \citep{Mehta22}, where the authors  investigated the depth dependence in QBOs by applying the methods of Empirical Mode Decomposition (EMD) and the FFT. They used oscillation frequencies from both GONG and MDI/Helioseismic and magnetic Imager instruments covering solar cycles 23 and 24, but did not find a clear depth dependence in the QBO periods in any of the cycles. Since all these modes travel to  the surface and reflect back from there, our findings of comparable QBO periods hint that the possible source lies in the near-surface shear layer (NSSL). Similar explanations for the location of QBO source region were also proposed in earlier studies, based on the helioseismic data \citep[e.g.,][]{Fletcher10, Broomhall12, Mehta22}.

\subsection{Is the second dynamo in NSSL responsible for QBOs?} \label{nssl}

\begin{figure}
\begin{center}
\includegraphics[width=0.98\textwidth,  height=5.15cm]{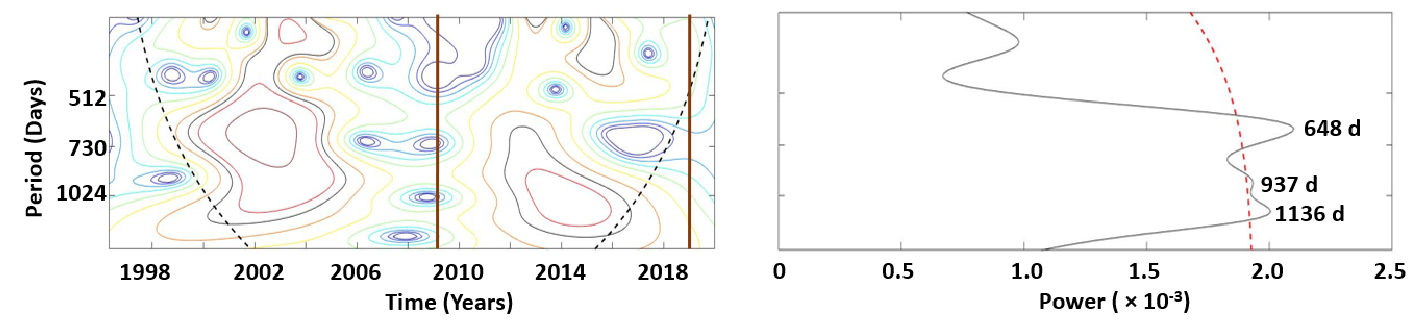}  
\caption{Morlet wavelet spectra (left) and the global power spectra (right) for all modes with lower-turning points in the upper 10\% layer below the surface.  The two vertical brown lines in left panels delineate the boundaries between cycles 23 and 24 then cycles 24 and 25,  while dashed lines in both panels represent the 95\% confidence level. }
\label{QBO_nssl}
\end{center}
\end{figure}

\begin{figure}
\begin{center}
\includegraphics[width=0.98\textwidth,  height=9.5cm]{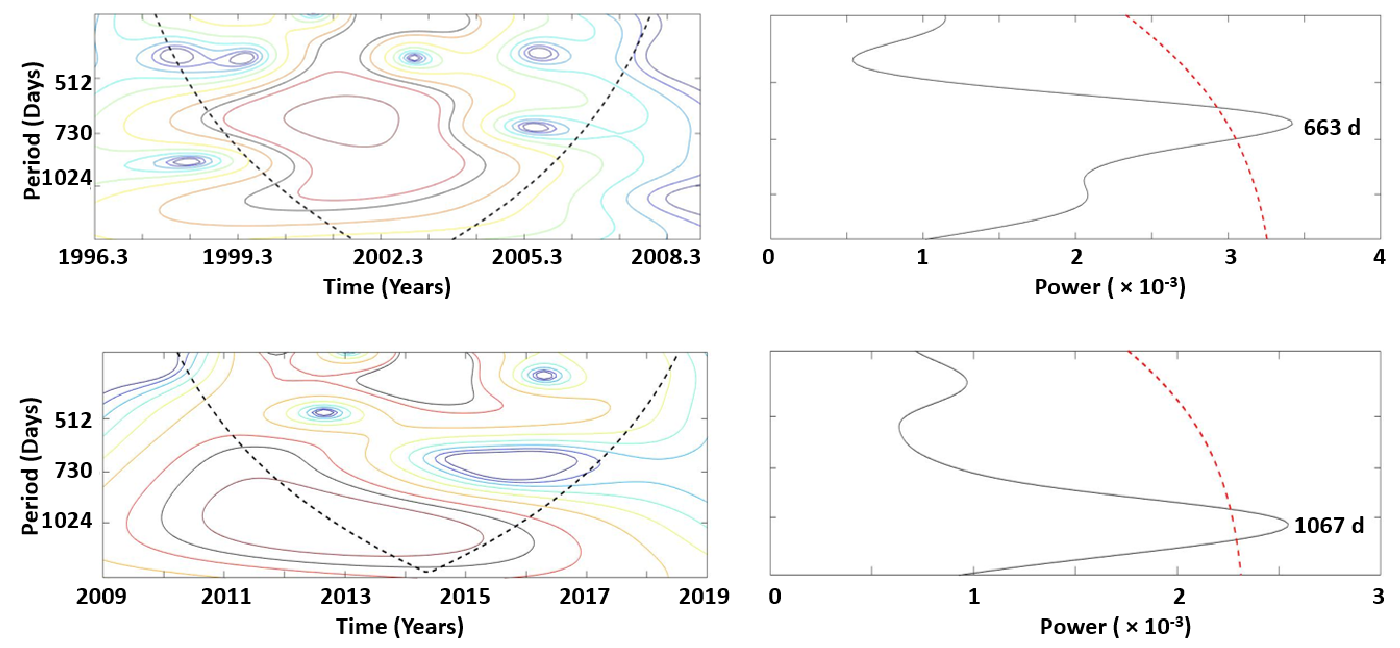}  
\caption{Morlet wavelet spectra (left) and the global power spectra (right) for all modes with lower-turning points in the upper 10\% layer below the surface in cycle 23 (top) and cycle 24 (bottom). The dashed lines represent the 95\% confidence level. 
\label{QBO_nssl_23_24_ind}}
\end{center}
\end{figure}

To examine the possibility of the QBO source location being in the NSSL,
we analyzed modes confined to this layer. It is conjectured that the second dynamo resides in the upper 5\% below the surface, thus we use  modes that have sensitivity to the properties of this layer only. Since a sufficient number of global acoustic modes in the upper 5\%  layer are currently not available, we widened the depth to 10\%.  This selection criterion automatically excludes modes that have lower-turning points below this layer.  The Morlet wavelet and global power spectra for all the modes confined to this layer are shown in Figure~\ref{QBO_nssl}  for the entire data set and in Figure~\ref{QBO_nssl_23_24_ind} for solar cycles 23 and 24 analyzed separately. We obtain two prominent QBO periods at 648$_{-56}^{+107} $  and  1136$_{-112}^{+57} $ days, with  one marginally significant peak at 937 days for the entire data and  peaks at 663$_{-68}^{+60}$ and 1067$_{-109}^{+125}$ days for cycles 23 and 24, respectively.   These QBO periods are consistent with those reported in previous sections for the modes probing deeper regions.  
Despite the fact that all modes spend maximum time near the surface, the modes restricted by their lower-turning points preclude any information below this turning-point radius. Since the analyses of modes carrying information from different regions below the NSSL provide similar QBO periods, we emphasize that a close agreement between the QBO periods in all cases is significantly influenced by the conditions in NSSL. If the changes in deeper layers were responsible for QBOs, we would not have obtained any QBO periods for the modes confined to the NSSL. Therefore, we suggest the plausible source location of QBOs to be within the near-surface layers.


\begin{figure}
\begin{center}
\includegraphics[width=0.98\textwidth,  height=14.5cm]{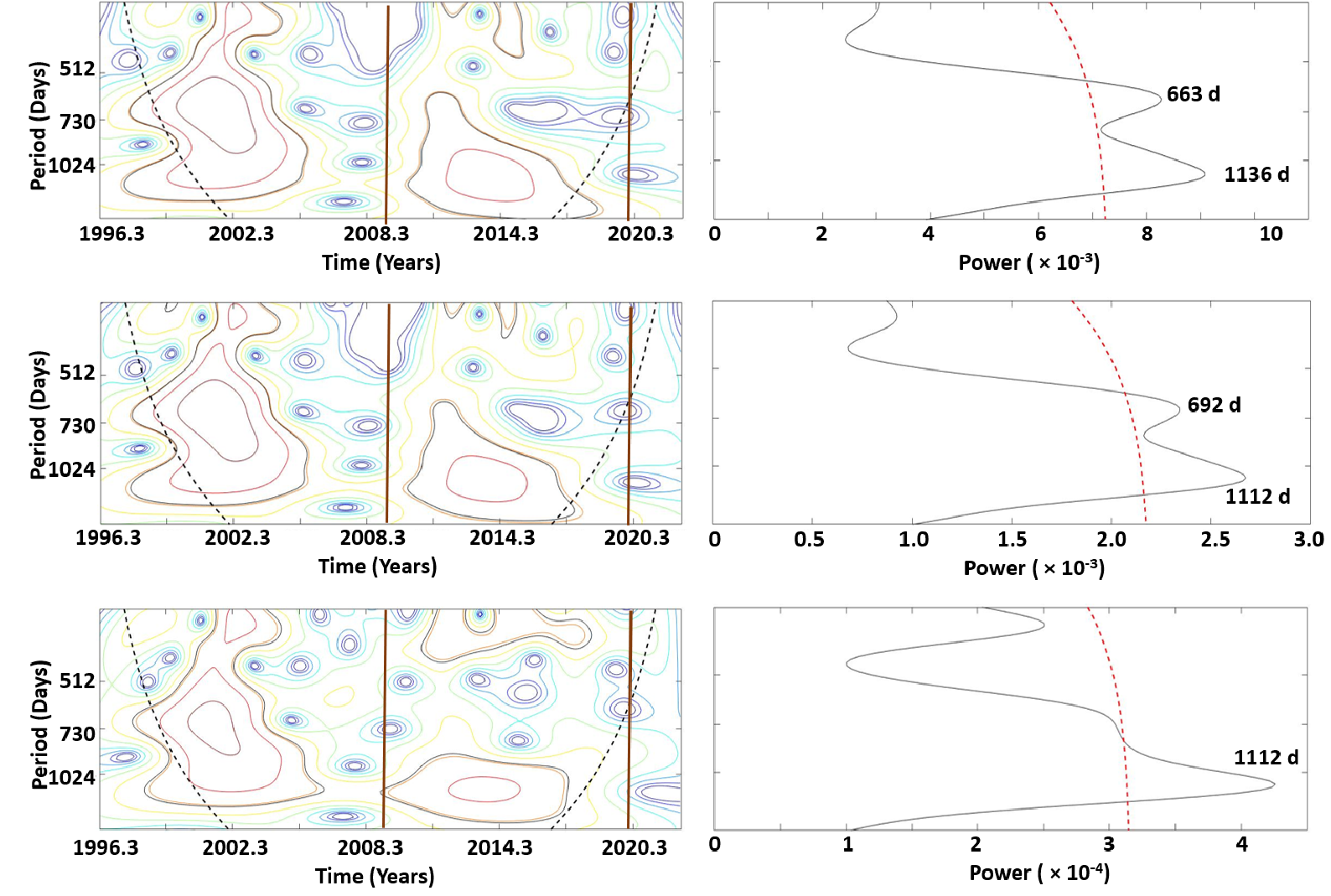}  
\caption{The same as Figure~\ref{QBO_depth_dep}  but for modes with upper-turning points in the high-$\nu$ (top), mid-$\nu$ (middle), and low-$\nu$ (bottom) bands. the dashed lines represent the 95\% confidence level.
\label{QBO_nu_dep}}
\end{center}
\end{figure} 

 We further study the influence of the changes in the NSSL on QBO periods. It was reported by \citet{Basu12} that there were differences in the properties of different oscillation modes caused by the changes in the Sun's shear layers during the declining phase of cycle 23. This was confirmed in subsequent studies by \citet{Howe17} and \citet{Jain18b}. Following \citet{Basu12},  we divided the entire data into  three frequency bands: 1860 $\le \nu \le$ 2400 $\mu$Hz (low-$\nu$), 2400 $< \nu \le$ 2920 $\mu$Hz (mid-$\nu$), and  2920 $< \nu \le$ 3450 $\mu$Hz (high-$\nu$).  It may be noted that for a given $\ell$, the upper-turning points lie closer to the surface with increasing frequencies. Since the mode characteristics of the higher-frequency range are also influenced by the properties of the modes corresponding to lower-frequency ranges, we emphasize that these frequency bands are not entirely independent.  We again utilized Equation~\ref{eqn_nu} and computed the weighted mean frequency shifts  for different frequency bands. The Morlet  and the GWPS  presented in  Figure~\ref{QBO_nu_dep} display similarities as well as differences between the different frequency ranges.  While we find a close agreement between the two QBO periods obtained for the mid-$\nu$ (692$^{+113}_{-84} $ and 1112$^{+157}_{-289} $ days) and high-$\nu$ bands (663$^{+142}_{-81} $ and 1136$^{+133}_{-277} $ days), there is only one peak in  the low-$\nu$ band representing the low-frequency part of the QBO spectrum with  a period of 1112$^{+157}_{-289} $ days. It is evident from   Figure~\ref{QBO_nu_dep}  that there was insufficient power buildup around 700 days  for the modes in  the low-$\nu$ band, which might be responsible for the single peak in this frequency band. To examine this, we repeated the analysis for cycles 23 and 24 separately in each frequency band. Similar to the earlier results, we obtain one QBO period corresponding to each frequency band in each cycle;  at 788$^{+380}_{-219} $, 692$^{+167}_{-122} $  and  692$^{+206}_{-120} $ days for cycle 23 in the low-$\nu$, mid-$\nu$ and high-$\nu$ frequency bands, respectively,  and at 1112$^{+336}_{-234} $, 1090$^{+261}_{-212} $  and  1112$^{+211}_{-234} $ days for cycle 24.   We notice that the QBO periods for all three frequency bands in cycle 24 are comparable. The peak period obtained for the low-$\nu$ band in cycle 23 is different from other two frequency bands, but  the differences are within the estimated uncertainties. We also find that the QBO power is strongest for the modes corresponding to the high-$\nu$ band, which  decreases by several orders for the lower-$\nu$ bands. This leads us to speculate that the QBO signals originate from a source lying in the layers where the power is stronger.  Since the selected frequency bands are not completely isolated, a precise location of the QBO source would require inverting global high-degree {\it p} modes that are not currently available.

\section{Discussion} \label{sec-discussion}

 We have presented a detailed analysis of the acoustic oscillation frequencies in order to explore short-term periodicities. While numerous studies have been carried out to characterize the periods shorter than the 11 yr cycle, such as the QBO type,  in the sunspots, the 10.7 cm radio flux and other measures of solar activity \citep[][and references therein]{cadavid05,Vecchio10,Chowdhury13,Chowdhury19,Deng16,Kilcik20,Chowdhury22,Ravindra22}, such studies based on the helioseismic data are comparatively fewer \citep{ Fletcher10,Broomhall12,Simoniello12, Simoniello13a,Inceoglu21, Inceoglu22,Mehta22}. Nonetheless, the QBO-like periodicities detected in solar activity and the oscillation data are more or less consistent with each other. Since the variations in helioseismic oscillation frequencies are sensitive to the changes in the solar interior, this similarity indicates that the surface activity measurements mostly reflect  the properties of the solar interior. However, it is important to note that the studies based on surface activity measurements do not provide the precise location of the QBO source, while  the helioseismic measurements have the unique ability to identify  the location of the source. This has been confirmed to some extent in several helioseismic studies.

\begin{figure}
\begin{center}
\includegraphics[width=0.98\textwidth,  height=5.15cm]{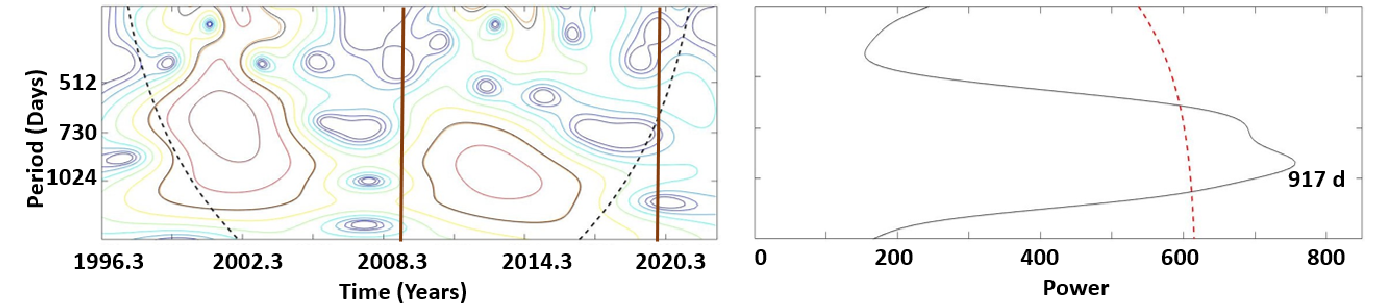}  
\caption{Morlet wavelet spectrum (left) and the global power spectrum (right) for 10.7 cm radio flux.  Daily flux values are averaged over 36 days. The two vertical brown lines in the left panels delineate the boundaries between cycles 23 and 24 then cycles 24 and 25,  while dashed lines in the right panels represent the 95\% confidence level.
\label{f10}}
\end{center}
\end{figure}

\begin{figure}
\begin{center}
\includegraphics[width=0.99\textwidth,  height=10.5cm]{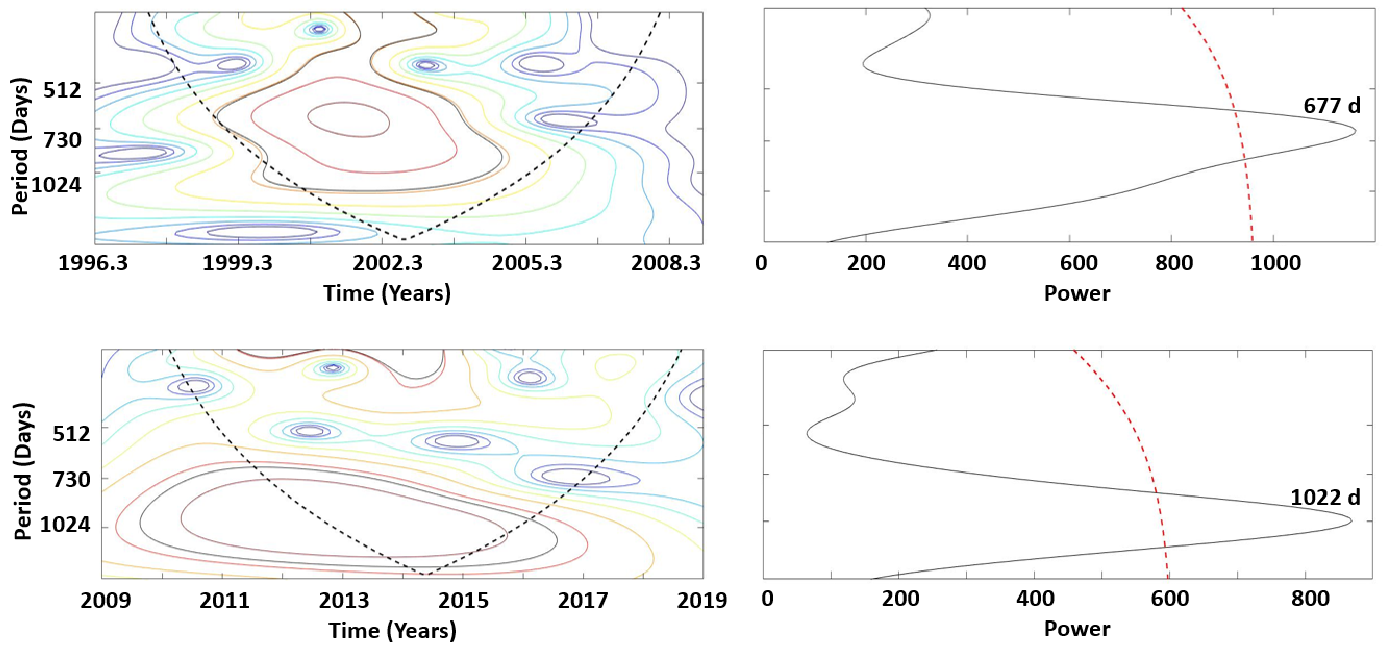}  
\caption{Morlet wavelet spectra (left) and the global power spectra (right) for 10.7 cm radio flux in cycle 23 (top) and cycle 24 (bottom). The dashed lines represent the 95\% confidence level.
\label{f10_ind}}
\end{center}
\end{figure}

 In this study based on the helioseismic data, we found differences in the QBO periods between cycles 23 and 24. Thus, it is of interest to explore if the differences in the QBO periods corresponding to different cycles are limited to the solar interior only or whetherthese are also present in the magnetic activity measured in the solar atmosphere.  For this purpose, we analyzed the daily 10.7 cm radio flux    measurements\footnote{https://www.spaceweather.gc.ca/forecast-prevision/solar-solaire/solarflux/sx-5-en.php} \citep{Tapping13}.  The 10.7 cm flux represents the
contributions from both strong (sunspots) and weak (radio plages) magnetic flux in the upper chromosphere, in addition to the quiet-Sun background emission. We averaged the radio flux values over the same 36 day time intervals as the frequency time series and analyzed them in a similar way as the frequency shifts.   As shown in Figure~\ref{f10}, we obtain two noticeable zones in the wavelet spectrum,  but only one dominant peak at 917$_{-296}^{+195} $ days in the global power spectrum in the analysis of the entire data. This contradicts the double-peak structure found in the oscillation mode frequencies discussed in Section~\ref{sec-results}. We speculate that this might have resulted from the lack of significant power with different prominent periods  in the QBO spectrum, and that the identified period is an average of quasiperiodic behavior spread over several periods.   The disparity between the QBO periods obtained for the radio flux and the acoustic modes suggests a complex relationship between the interior and atmosphere, and has been discussed by several authors \citep[e.g.,][]{Tripathy07,Cristina19}.    It is intriguing to note from Figure~\ref{f10_ind} that the separate analyses for both cycles confirm two different QBO periods:   677$_{-82}^{+164} $ and 1022$_{-181}^{+196} $ days for cycles 23 and 24, respectively. These are  comparable with the QBO periods found in our analysis of acoustic modes. 
 
To understand the origin of QBOs and their periods, \citet{Benevolenskaya98} proposed a model based on the idea of two dynamo sources separated in space;  the first source is located near the bottom of the convection zone, with the second source operating in the NSSL. Though the dynamo theories successfully explain the 11 yr cycle to some extent, the significantly different  predictions for cycle 24  based on different dynamo theories raise some questions about our understanding  of the solar dynamo \citep{Pesnell12}. In addition, the quasi-biennial cycle poses another challenge to these theories.   Our study suggests the possible origin of QBOs in a layer closer to the surface, while a few earlier studies suggest that these are generated by an interaction between two oppositely signed magnetic activity bands seated in the deep interior of the opposite hemispheres. The later perception appears to be supported by  the recent numerical simulations of \citet{Veronig21}, where authors demonstrated  that the solar cycle evolves independently in the two hemispheres.  

 Other suggested mechanisms responsible for QBOs include the instability of the magnetic Rossby waves in the tachocline \citep{Zaqarashvili10}, periodic energy exchange between the Rossby waves, differential rotation, and the toroidal field via tachocline nonlinear oscillations \citep{Dikpati18}, and the interplay between the flow and magnetic fields, where the turbulent $\alpha$-mechanism works in the lower half of the solar convection zone and extends to the surface \citep{Inceoglu19}.  In a subsequent study, \citet{Inceoglu22} argued that the source region of QBOs is below 0.78~R$_{\Sun}$. \citet{Zaqarashvili18} studied the influence of a toroidal magnetic field on the dynamics of shallow water waves in the solar tachocline. The author found that the toroidal magnetic field splits equatorial Rossby and Rossby-gravity waves into fast and slow modes. While the global equatorial fast magneto-Rossby waves have a periodicity of 11 yr, matching the timescale of activity cycles with the solutions confined around sunspot activity belts, the equatorial slow magneto-Rossby-gravity waves have the periodicity of 1--2 yr, which may correspond to observed annual oscillations and QBOs. Evidence of the existence of equatorial Rossby waves has been presented in several studies using the helioseismic data \citep[e.g.,][]{Loptien18,Liang19,Hanson20}.  In addition, \citet{McIntosh17} have also detected the presence of  Rossby-type waves in coronal bright points obtained by the Extreme-Ultraviolet Imager  instruments on the Solar Terrestrial Relations Observatory spacecraft, and the Atmospheric  Imaging Assembly  instrument on the  Solar Dynamic Observatory  spacecraft. These waves propagate in the retrograde direction relative to the rotation and have been studied in detail by several authors \citep[see the review by][]{Zaqarashvili21}.

On the other hand, some studies suggest that QBOs might originate from the dynamo action in the layers just below the surface where the second dynamo is situated. \citet{Strugarek18} performed a series of 3D nonlinear MHD simulations by varying the rotation rate and luminosity of the modeled solar-like convective envelopes. They   found that the shorter cycles,  located at the top of the convective envelope close to the equator,  are observed in numerical experiments for the small values of the local Rossby number,  while a moderate  Rossby number  is needed  for the decadal magnetic cycles originating near the base of the convection zone. The deep-seated dynamo sustained in these numerical experiments is fundamentally nonlinear, thus it is the feedback of the large-scale magnetic field on the differential rotation that sets the magnetic cycle period. The 
authors also found that the cycle period decreases with the Rossby number, which offers an alternative theoretical explanation for the variety of activity cycles observed in solar-like stars. 

In addition, using low-degree frequencies from Birmingham Solar Oscillation Network, \citet{Broomhall12} interpreted the seismic signatures of the QBOs as a result of the second dynamo mechanism seated near the bottom of the layer extending 5\% below the solar surface.    Recently, \citet{Mehta22} supported these findings, stating that the magnetic field responsible for producing QBOs in the frequency shifts of {\it p}-modes is anchored above approximately the upper 5\% of the solar interior.  They also found that the presence of the QBOs is not sensitive to the depth to which the {\it p}-mode traveled, nor to the average frequency of the {\it p}-mode. \citet{Simoniello13a} have discussed a different scenario and suggested that the observed properties could result from the beating between a dipole and quadrupole magnetic configuration of the dynamo. The understanding of the mechanisms responsible for QBOs has been advanced significantly in recent years, but it  is not yet fully understood.

\section{Summary} \label{sec-summary}

In summary, by analyzing the acoustic mode oscillation frequencies from GONG for cycles 23 and 24, we affirm that the QBO-type signals are present in both cycles. The amplitudes are found to vary with the progression of the cycle, appearing higher during the high-activity phase, with a subsequent decrease during the minimum-activity period. This is not unusual, as similar results have been reported in earlier studies \citep[e.g.,][]{Kolotkov15}. However, the most striking features found in this study are the double-peak structure in the global wavelet spectra and different QBO periods in cycles 23 and 24. The dominant QBO periods are found to be about 2 and 3 years in cycles 23 and 24, respectively.  Since these periods are found to be influenced by the changes in the near-surface layers, we conjecture that their source might be located in these layers.  Note that  consistent helioseismic data are currently available only for two solar cycles having different characteristics, thus continuous measurements for several solar cycles and their inclusion in solar dynamo models are required for a better understanding of the origin of QBOs.

\begin{acknowledgments}
We thank the reviewer for several useful suggestions. This work utilizes GONG data obtained by the NSO Integrated Synoptic Program, managed by the National Solar Observatory, which is operated by the Association of Universities for Research in Astronomy (AURA), Inc. under a cooperative agreement with the National Science Foundation and with a contribution from the National Oceanic and Atmospheric Administration. The GONG network of instruments is hosted by the Big Bear Solar Observatory, High Altitude Observatory, Learmonth Solar Observatory, Udaipur Solar Observatory, Instituto de Astrof\'{\i}sica de Canarias, and Cerro Tololo Inter-American Observatory. K.J. and S.C.T. acknowledge partial funding from the NASA DRIVE Science Center COFFIES Phase II grant 80NSSC22M0162 to Stanford University.
\end{acknowledgments}

\facilities{GONG}

\bibliography{Jain_ms}{}
\bibliographystyle{aasjournal}

\end{document}